\documentclass[runningheads]{llncs}
\usepackage[T1]{fontenc}
\usepackage{graphicx}
\usepackage[most]{tcolorbox}
\usepackage{xcolor}
\usepackage{booktabs}
\usepackage{tabularx}
\usepackage{subcaption}
\usepackage{amsmath,amssymb}

\usepackage[ruled,vlined]{algorithm2e}

\begin{document}
\title{Synergizing a Decentralized Framework with LLM-Assisted Skill and Willingness-Aware Task Assignment for Volunteer Crowdsourcing}
\titlerunning{Hybrid VCS Framework with LLM and Blockchain}
%
\author{Riya Samanta\inst{1}\orcidID{0000-0002-8156-7636} \and
Rituparna Bhattyacharya\inst{1}\orcidID{0009-0008-2746-3693}}

\authorrunning{Riya et al.}

\institute{Techno India University, Salt Lake, West Bengal, India\\
\email{riya.s.rituparna.b@technoindiaeducation.com}}
\maketitle              
\begin{abstract}
Volunteer crowdsourcing (VCS) platforms increasingly support education, healthcare, disaster response, and smart city applications, yet assigning volunteers to complex tasks remains challenging due to fine-grained skill heterogeneity, unstructured profiles, dynamic willingness, and bursty workloads. Existing methods often rely on coarse or keyword-based skill representations, resulting in poor matching quality. We propose a hybrid VCS framework that integrates LLM-assisted semantic preprocessing, an interpretable skill- and willingness-aware assignment engine, and blockchain-enforced execution. The LLM is used only to extract and canonicalize fine-grained skills and preference cues from unstructured resumes and task descriptions, while assignment is performed by a utility-driven matcher that models partial skill coverage and participation likelihood. Smart contracts provide transparent and tamper-resistant enforcement without on-chain optimization overhead. Experiments on diverse resume datasets show a 42.3\% improvement in assignment utility over skill-only greedy matching and an increase in task coverage from 0.80 to 0.90. These results highlight the value of combining semantic intelligence, interpretable matching, and decentralized enforcement for effective volunteer-task allocation.

\keywords{Volunteer Crowdsourcing \and Skill-Oriented Task Assignment \and Willingness-Aware Matching \and LLM-Assisted Semantic Normalization \and Blockchain-Enforced Decentralized Systems}
\end{abstract}

\section{Introduction}
\vspace{-0.1in}
Volunteer crowdsourcing (VCS) has emerged as an effective paradigm for addressing societal challenges by mobilizing individuals’ skills and expertise toward tasks of social importance~\cite{Sama2212:Volunteer}. Modern VCS platforms support applications in education, public health, disaster response, environmental monitoring, and smart city services under dynamic participation and resource constraints. Unlike microtask-oriented crowdsourcing, many VCS tasks are complex and skill-intensive, requiring fine-grained competencies, domain expertise, and contextual awareness~\cite{samanta2021swill,Cheng2016,Chen2014}. As a result, the effectiveness of a VCS platform largely depends on accurate volunteer–task assignment under heterogeneous skills and participation behavior. A central challenge is \emph{skill-oriented task assignment under unstructured information}. Volunteer capabilities and task requirements are typically expressed in free-form text such as resumes, self-descriptions, or activity logs~\cite{pazzani2007content}. Existing methods often rely on keyword matching or coarse skill taxonomies, which fail to capture semantic equivalence, contextual relevance, and partial skill coverage, leading to inefficient volunteer utilization~\cite{borgave2023resume}.

Beyond skill compatibility, \emph{volunteer willingness} strongly influences task outcomes. Even when volunteers possess the required expertise, factors such as interest, availability, and perceived effort affect whether tasks are accepted and completed. Prior work shows that ignoring willingness increases dropout rates and reduces platform efficiency~\cite{samanta2021swill,10925570,samanta2024empowering,samanta2024sustainable}. Effective assignment must therefore jointly model skill suitability and participation likelihood. Recent advances in large language models (LLMs) provide a practical mechanism for extracting structured signals such as skills, experience, and preference cues
from unstructured text. Meanwhile, blockchain-based decentralized applications improve transparency and trust through immutable records and smart contracts. However, existing crowdsourcing systems rarely integrate semantic preprocessing with interpretable task assignment under decentralized execution.

To address these challenges, we present \textbf{SWATi}, a skill- and willingness-aware task assignment framework for volunteer crowdsourcing that combines LLM-assisted semantic preprocessing, interpretable utility-driven matching, and blockchain-enforced execution. LLMs extract and normalize skills and preference cues from unstructured profiles and task descriptions, while the assignment engine models partial skill coverage and time-varying willingness to produce transparent allocation decisions. Smart contracts then enforce finalized assignments and immutably record task states.

\paragraph{Contributions.}
The main contributions of this work are:
\vspace{-0.05in}
\begin{itemize}
    \item \textbf{LLM-based semantic preprocessing} for extracting and normalizing skills and preference cues from unstructured volunteer and task descriptions.
    \item \textbf{SWATi}, a utility-driven task assignment framework that jointly models fine-grained skill compatibility and volunteer willingness.
    \item A \textbf{blockchain-backed execution layer} using smart contracts for transparent and tamper-resistant assignment realization.
    \item \textbf{Empirical evaluation} on heterogeneous resume datasets demonstrating improved matching quality and platform utility.
\end{itemize}

\section{Related Work}
\vspace{-0.1in}

Early crowdsourcing task assignment mainly targets micro-task settings with independent, low-skill tasks, making such models inadequate for volunteer crowdsourcing services (VCS), where tasks are often complex and skill-intensive. As a result, prior work has studied expertise-aware selection~\cite{goel2014}, multi-skill and spatial assignment~\cite{Cheng2016,Liu2016}, and dynamic formulations with real-time arrivals, task dependencies, and willingness-aware participation~\cite{Song2020,Ni2020,samanta2021swill,samanta2025enhancing}. Other studies incorporate collaboration, proximity, and disaster-response constraints~\cite{samantafogirecruiter,Sama2212:Volunteer,10925570}. However, these approaches largely assume structured inputs with explicit skill labels. Blockchain has also been explored for decentralized crowdsourcing because of its transparency, immutability, and smart-contract-based automation~\cite{nakamoto2008bitcoin,wood2014ethereum}. Systems such as CrowdBC~\cite{li2018crowdbc}, MCS-Chain~\cite{feng2019mcs}, MetaCrowd~\cite{le2024metacrowd}, Cerebrum~\cite{cerebrum2025decentralized}, Golem~\cite{golem2016decentralized}, and Filecoin~\cite{filecoin2017} demonstrate decentralized execution and verification, but generally treat assignment as a system function rather than a semantically grounded matching problem. More recent blockchain-based assignment approaches consider preferences, reputation, and distributed decision making~\cite{xu2022blockchain,10126111,li2022blockchain,9874795}, yet still rely on structured worker and task representations. Meanwhile, large language models (LLMs) have shown strong ability to extract structured signals such as skills and experience from unstructured text~\cite{brown2020language,ouyang2022training}. However, their use as semantic preprocessing modules for volunteer crowdsourcing task assignment remains largely unexplored.

\textit{To the best of our knowledge, no prior work jointly integrates LLM-based semantic skill normalization, skill- and willingness-aware utility-driven assignment, and blockchain-enforced execution within a unified VCS framework.}

\begin{figure*}[t]
    \centering
    \includegraphics[width=0.9\linewidth]{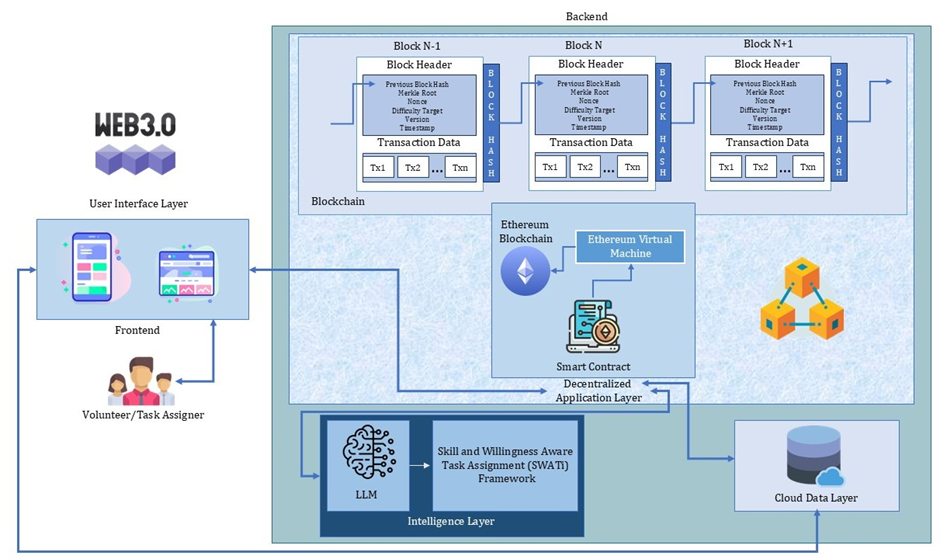}
    \footnotesize
    \caption{System architecture of the proposed framework: LLM-based semantic preprocessing, algorithmic task assignment, and blockchain-enforced execution.}
    \label{fig:architecture}
    \vspace{-0.2in}
\end{figure*}

\vspace{-0.1in}
\section{Proposed Framework}
\vspace{-0.1in}
We consider a volunteer crowdsourcing (VCS) platform operating under dynamic task and volunteer arrivals. Let $\mathcal{V}=\{v_1,\dots,v_N\}$ denote the active volunteers and $\mathcal{T}=\{t_1,\dots,t_M\}$ the pending tasks at a decision epoch. Each task $t_j\in\mathcal{T}$ requires a skill set $\mathcal{S}_{t_j}$, while each volunteer $v_i\in\mathcal{V}$ possesses a skill set $\mathcal{S}_{v_i}$. In practice, however, these sets are not explicitly available and must be inferred from unstructured sources such as task descriptions, resumes, profiles, or prior interaction records.

\vspace{-0.08in}
\subsection{LLM-Based Semantic Skill Normalization}
\label{subsec:llm_skill_norm}

Let $D_{v_i}$ and $D_{t_j}$ denote the textual descriptions of volunteer $v_i$ and task $t_j$, respectively. We use a large language model (LLM) as a \emph{semantic normalization layer} that transforms heterogeneous free-form text into structured representations~\cite{brown2020language,ouyang2022training,hao2022structured}. Formally, an extractor $\Phi_\theta(\cdot):\mathcal{D}\!\rightarrow\!\mathcal{Z}$ maps text into a schema containing normalized skills, evidence spans, proficiency indicators, and preference cues. Unlike keyword matching, $\Phi_\theta(\cdot)$ captures synonymy and compositional skills; for example, ``computer vision using YOLOv8'' may be normalized to \{Computer Vision, Object Detection, YOLO\}.

Given tokenized input $\mathbf{x}=(x_1,\dots,x_L)$, the transformer computes contextual representations as $\mathbf{H}^{(\ell)}=\mathrm{TransformerBlock}(\mathbf{H}^{(\ell-1)})$, with $\mathbf{H}^{(0)}=\mathrm{Embed}(\mathbf{x})$, and attention $\mathrm{Attn}(\mathbf{Q},\mathbf{K},\mathbf{V})=\mathrm{softmax}\!\left(\frac{\mathbf{Q}\mathbf{K}^\top}{\sqrt{d_k}}\right)\mathbf{V}$. Under a structured output schema (e.g., JSON), the model generates tuples $(s,\texttt{evidence},\rho,\pi)$ denoting canonical skills, supporting spans, proficiency scores, and optional willingness cues.

The extracted skills are aligned to a controlled vocabulary $\mathcal{S}$ through an alias-resolution operator $\mathcal{A}(\cdot)$, yielding $\mathcal{S}_{v_i}=\mathcal{A}(\Phi_\theta(D_{v_i}))$ and $\mathcal{S}_{t_j}=\mathcal{A}(\Phi_\theta(D_{t_j}))$. This step merges lexical variants, enforces type consistency, and optionally supports hierarchical roll-up. In addition, semantic embeddings $\mathbf{e}(D)\in\mathbb{R}^d$ are retained to capture implicit information, giving hybrid representations $\mathcal{R}_{v_i}=(\mathcal{S}_{v_i},\mathbf{e}(D_{v_i}))$ and $\mathcal{R}_{t_j}=(\mathcal{S}_{t_j},\mathbf{e}(D_{t_j}))$. To improve reliability, the extractor operates under schema constraints, controlled decoding, and evidence citation; an optional verifier rejects malformed outputs. The LLM is used only for semantic extraction, while assignment remains fully algorithmic.

\vspace{-0.08in}
\subsection{Skill, Content, and Willingness Modeling}
\label{subsec:similarity_willingness}

Given the canonicalized skill sets, volunteer--task compatibility is measured using explicit skill overlap and contextual similarity. First, we define set-based skill similarity using Jaccard overlap~\cite{cazzanti2006information} as
$\text{SkillSim}(v_i,t_j)=\frac{|\mathcal{S}_{v_i}\cap\mathcal{S}_{t_j}|}{|\mathcal{S}_{v_i}\cup\mathcal{S}_{t_j}|}$,
which captures partial rather than binary feasibility. Second, to model implicit alignment, we compute TF-IDF vectors $\mathbf{x}_{v_i}$ and $\mathbf{x}_{t_j}$ from $D_{v_i}$ and $D_{t_j}$ and define cosine-based content similarity~\cite{deerwester1990indexing,jaccard1901distribution} as
$\text{ContentSim}(v_i,t_j)=\frac{\mathbf{x}_{v_i}\cdot \mathbf{x}_{t_j}}{\|\mathbf{x}_{v_i}\|\,\|\mathbf{x}_{t_j}\|}$.
Together, these two measures balance interpretability and robustness to underspecified requirements.

Beyond compatibility, effective assignment also depends on a volunteer’s willingness to accept and complete a task. We model willingness as a task-dependent probability $w_{i,j}(\tau)\in[0,1]$, where $w_{i,j}(\tau)=\mathbb{P}(A_{i,j}=1\mid \mathcal{C}_{i,j}(\tau))$, $A_{i,j}\in\{0,1\}$ denotes acceptance, and $\mathcal{C}_{i,j}(\tau)$ denotes contextual features. Since static resume datasets do not provide explicit acceptance labels, willingness is inferred from interpretable preference cues extracted during semantic preprocessing, such as domain affinity, prior exposure, stated interests, volunteering history, and availability indicators. For each pair $(v_i,t_j)$, these cues form a low-dimensional vector $\mathbf{p}_{i,j}$, which is mapped by a deterministic scoring function $f(\cdot)$. When participation history $\mathcal{H}_i$ is available, a task-conditioned tendency $g(\mathcal{H}_i,t_j)$ is also incorporated. The final estimate is
$w_{i,j}(\tau)=\sigma\!\left(\eta\,g(\mathcal{H}_i,t_j)+(1-\eta)\,f(\mathbf{p}_{i,j})\right)$,
where $\eta\in[0,1]$ balances history and profile-derived cues, and $\sigma(\cdot)$ normalizes the score. To capture temporal variation, willingness is updated by exponential smoothing:
$w_{i,j}(\tau)=\lambda\,w_{i,j}(\tau^-)+(1-\lambda)\,\hat{w}_{i,j}(\tau)$,
where $\lambda\in[0,1]$ controls responsiveness. Thus, LLMs are used only to extract preference signals; willingness estimation itself remains interpretable and algorithmic.

\vspace{-0.08in}
\subsection{Utility-Based Assignment Objective}
\label{subsec:utility-objective}

To jointly capture skill suitability, contextual alignment, and participation likelihood, we define the utility of assigning volunteer $v_i$ to task $t_j$ at time $\tau$ as
$U(v_i,t_j,\tau)=\alpha\,\text{SkillSim}(v_i,t_j)+\beta\,\text{ContentSim}(v_i,t_j)\,w_{i,j}(\tau)$,
where $\alpha,\beta\in[0,1]$ and $\alpha+\beta=1$. Here, willingness acts as a multiplicative feasibility factor that discounts otherwise compatible assignments when likely participation is low.

The assignment problem is then to select a matching $A(\tau)\subseteq \mathcal{V}\times\mathcal{T}$ that maximizes the aggregate expected utility:
$\max_{A(\tau)} \sum_{(v_i,t_j)\in A(\tau)} U(v_i,t_j,\tau)$,
subject to platform constraints such as volunteer capacity, deadlines, availability windows, and fairness or policy requirements.

\vspace{-0.05in}
\subsection{Dataset Description}
\label{subsec:dataset}
\vspace{-0.1in}

We construct a heterogeneous resume corpus from \textbf{three complementary sources} to capture the structural noise and semantic variability typical in volunteer crowdsourcing platforms. The primary dataset is a \textbf{Kaggle resume corpus}~\cite{snehaanbhawal_resume_dataset} containing over \textbf{2,400 resumes} across \textbf{24 occupational domains} in PDF, text, and HTML formats. Due to high structural noise and non-canonical skill expressions, we restrict our analysis to the \textbf{engineering stream}, yielding \textbf{118 resumes} related to Computer Science and IT roles. To complement this, we generate a controlled \textbf{synthetic subset} of \textbf{120 resumes} using \textbf{GPT-5.2} with a fixed prompt, preserving realistic linguistic variation while enforcing consistent structure and skill coverage. This combination enables evaluation of semantic extraction under both noisy real-world and controlled settings.

\begin{algorithm}[t]
\footnotesize
\caption{SWATi}
\label{alg:swati-greedy}
\DontPrintSemicolon
\SetKwInOut{Input}{Input}
\SetKwInOut{Output}{Output}
\SetKwFunction{LLM}{LLMExtract}
\SetKwFunction{Alias}{AliasResolve}
\SetKwFunction{TFIDF}{TFIDF}
\SetKwFunction{Cos}{CosSim}
\SetKwFunction{Jacc}{Jaccard}
\SetKwFunction{Sig}{Sigmoid}

\Input{$V=\{v_i\}$ with $D_{v_i}$; $T=\{t_j\}$ with $D_{t_j}$; skill ontology $\mathcal{S}$; history $\{H_i\}$; weights $\alpha,\beta$ ($\alpha+\beta=1$); mixing $\eta$; smoothing $\lambda$; capacities $\{c_i\}$}
\Output{$A(\tau)\subseteq V\times T$}

\ForEach{$v_i\in V$}{
$S_{v_i}\leftarrow \Alias(\LLM(D_{v_i}),\mathcal{S})$; $x_{v_i}\leftarrow \TFIDF(D_{v_i})$; $load(v_i)\leftarrow0$\;
}
\ForEach{$t_j\in T$}{
$S_{t_j}\leftarrow \Alias(\LLM(D_{t_j}),\mathcal{S})$; $x_{t_j}\leftarrow \TFIDF(D_{t_j})$\;
}

\ForEach{$(v_i,t_j)\in V\times T$}{
$s\leftarrow \Jacc(S_{v_i},S_{t_j})$; 
$c\leftarrow \Cos(x_{v_i},x_{t_j})$\;
$\hat{w}\leftarrow \Sig(\eta g(H_i,t_j)+(1-\eta)f(p_{i,j}))$\;
$w\leftarrow \lambda w_{i,j}(\tau^{-})+(1-\lambda)\hat{w}$\;
$U_{i,j}\leftarrow(\alpha s+\beta c)w$\;
}

$A(\tau)\leftarrow\emptyset$; sort $(v_i,t_j)$ by $U_{i,j}$ descending\;
\ForEach{$(v_i,t_j)$}{
\If{$t_j$ unassigned $\land load(v_i)<c_i$}{
$A(\tau)\leftarrow A(\tau)\cup\{(v_i,t_j)\}$; $load(v_i)++$\;
}
}
\Return{$A(\tau)$}
\end{algorithm}
\vspace{-0.1in}

\subsection{Skill- and Willingness-aware Task Assignment (SWATi)}
\vspace{-0.1in}
We now describe the execution flow of the proposed skill- and willingness-aware task assignment algorithm, \textbf{SWATi}. SWATi (Algorithm~\ref{alg:swati-greedy}) implements the utility-based assignment objective defined in Section~2.4 using a deterministic greedy procedure. At each decision epoch $\tau$, the algorithm considers all feasible volunteer--task pairs $(v_i,t_j)$ with associated utilities $U(v_i,t_j,\tau)$, assuming semantic normalization, similarity computation, and willingness estimation have already been performed. First, utilities are computed for all feasible pairs and sorted in non-increasing order. This ordering induces a priority sequence over candidate assignments and enables comparison across heterogeneous volunteer--task combinations. The algorithm then iterates through the ordered list. A volunteer $v_i$ is assigned to task $t_j$ if the task is unassigned and the capacity constraint of $v_i$ is satisfied. Once assigned, task $t_j$ is removed from further consideration and the load of $v_i$ is updated. The procedure terminates when no additional feasible assignments remain. This greedy strategy provides an efficient approximation to the utility maximization objective while preserving determinism and interpretability. Given identical inputs at epoch $\tau$, the algorithm produces identical assignments, supporting reproducible and auditable execution. The reliance on simple sorting and capacity checks also makes the method suitable for constrained environments, including blockchain-based smart contracts.

\vspace{-0.1in}
\subsection{Blockchain-Enforced Deployment}
\vspace{-0.1in}
We deploy \textbf{SWATi} using a hybrid on-chain/off-chain architecture (Figure~\ref{fig:architecture}) that separates semantic processing and optimization from decentralized enforcement. The system consists of four layers: \textit{User Interface}, \textit{Decentralized Application (DApp)}, \textit{Intelligence}, and \textit{Cloud Data}. \textbf{Ethereum}~\cite{kushwaha2022ethereum} serves as the blockchain substrate, where smart contracts enforce assignment states, task transitions, and immutable audit logs. The \textbf{User Interface Layer} allows volunteers and task publishers to register, submit tasks, and view assignments through a simple interface, hiding the complexity of blockchain interactions. The \textbf{DApp Layer} manages the lifecycle of tasks and records finalized task–volunteer assignments on-chain. Smart contracts maintain the integrity of assignment states and provide transparent, tamper-resistant records that support accountability and dispute resolution in decentralized volunteer ecosystems. The \textbf{Intelligence Layer} operates entirely off-chain and integrates an LLM-based semantic preprocessing module with the \textbf{SWATi} matching engine. The LLM extracts structured skill and preference signals from unstructured volunteer profiles and task descriptions, while SWATi performs deterministic utility-based matching using skill similarity, content relevance, and willingness. Executing this stage off-chain preserves scalability and reduces blockchain execution costs. Finally, the \textbf{Cloud Data Layer} stores rich off-chain data, including volunteer profiles, task descriptions, extracted skills, and historical records. Only authoritative assignment states are stored on-chain, enabling scalable matching while preserving transparency and trust.

\section{Preliminary Results}
We present preliminary results to evaluate the effectiveness and system-level trade-offs of \textbf{SWATi} using normalized utility, task coverage, execution time, and semantic extraction statistics on the unified resume dataset.

\begin{table}[!t]
\centering
\footnotesize
\caption{Semantic skill extraction statistics across LLM backends}
\label{tab:skill_extraction_full}
\setlength{\tabcolsep}{3pt}
\renewcommand{\arraystretch}{0.8}
\begin{tabularx}{\columnwidth}{lccc}
\toprule
\textbf{LLM Backend} & \textbf{Total Skills} & \textbf{Unique Vocabulary} & \textbf{Avg./Resume} \\
\midrule
Qwen3-Coder-480B-Cloud & 4234 & 170 & 12 \\
DeepSeek-V3 & 4528 & 178 & 13 \\
Qwen2.5-Coder-32B & 3812 & 160 & 11 \\
\bottomrule
\end{tabularx}
\vspace{-0.2in}
\end{table}

\paragraph{1. Impact of LLM-Based Semantic Skill Extraction.}
Table~\ref{tab:skill_extraction_full} compares skill extraction across LLM backends. Larger models such as DeepSeek-V3 extract more skills per resume, indicating higher recall for implicit competencies. However, after alias resolution and canonicalization, these gains do not translate into proportional improvements in assignment utility. We therefore select \textbf{Qwen3-Coder-480B-Cloud} as the primary extractor, as it offers a better trade-off among semantic coverage, latency, and cost. This highlights that SWATi’s gains arise mainly from its integrated matching design rather than dependence on a specific LLM.

\begin{figure*}[!t]
    \centering
    \begin{subfigure}{0.40\textwidth}
        \centering
        \includegraphics[width=\linewidth]{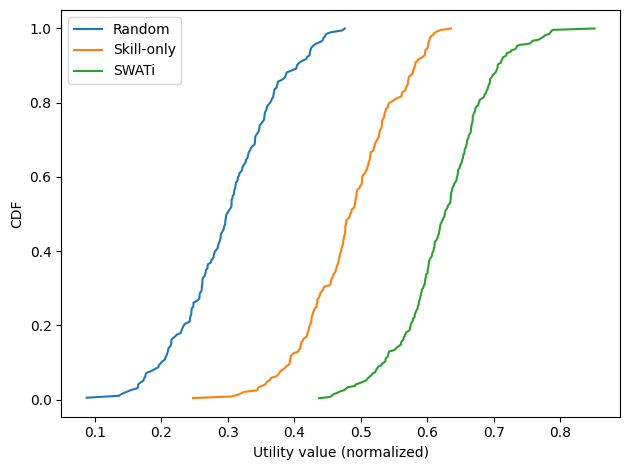}
        \caption{CDF of utility distribution}
        \label{fig:cdf_utility}
    \end{subfigure}
    \hfill
    \begin{subfigure}{0.41\textwidth}
        \centering
        \includegraphics[width=\linewidth]{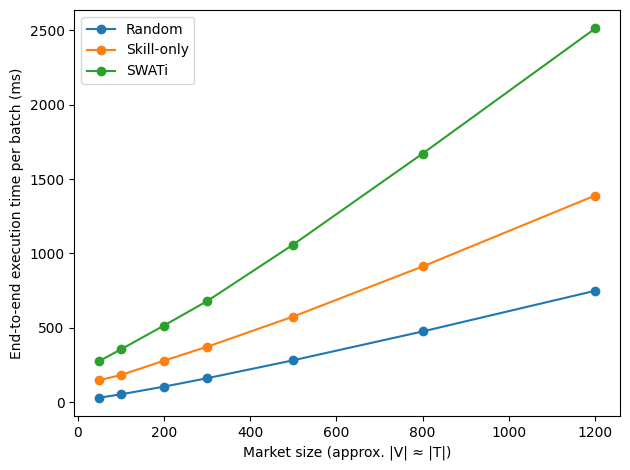}
        \caption{End-to-end execution time}
        \label{fig:execution_time}
        \vspace{-0.1in}
    \end{subfigure}
     \footnotesize
    \caption{Assignment quality and system-level evaluation on the unified resume dataset  $N{=}342$) with $|T|{=}300$ tasks.}
    \label{fig:evaluation_overview}
    \vspace{-0.2in}
\end{figure*}

\paragraph{2. Assignment Quality.}
Table~\ref{tab:assignment_quality} and Fig.~\ref{fig:evaluation_overview}(a) show that SWATi consistently outperforms Random and Skill-only baselines on the unified dataset ($N=342$, $T=300$). SWATi improves total utility by 42.3\% over Skill-only matching (167.4 vs.\ 117.6) while increasing task coverage from 0.80 to 0.90. The CDF in Fig.~\ref{fig:evaluation_overview}(a) further shows that SWATi yields higher utility across the full distribution, not just for a small subset of assignments. These gains come from jointly modeling skill compatibility, semantic alignment, and willingness.

\paragraph{3. Execution Time and Trade-offs.}
Fig.~\ref{fig:evaluation_overview}(b) reports batch execution time as market size increases ($|V|\approx|T|$). Random assignment is fastest, followed by Skill-only matching, while SWATi incurs additional overhead due to semantic preprocessing and willingness estimation. Nevertheless, SWATi scales smoothly without superlinear growth, indicating that improved match quality is achieved without sacrificing practical deployability. For volunteer crowdsourcing platforms, this trade-off is acceptable because assignment quality and participation likelihood are typically more critical than minimal per-batch latency~\cite{samanta2024sustainable}.

\begin{table}[h]
\vspace{-0.2in}
\centering
\footnotesize
\caption{Assignment quality comparison using normalized utility $U\in[0,1]$ on $|T|=300$ tasks evaluated over the unified resume dataset ($N=342$).}
\setlength{\tabcolsep}{3pt}
\renewcommand{\arraystretch}{0.7}
\label{tab:assignment_quality}
\begin{tabularx}{\columnwidth}{lccc}
\toprule
\textbf{Method} & \textbf{Total utility} & \textbf{Avg. utility} & \textbf{Task coverage} \\
\midrule
Random assignment   & 58.5  & 0.30 & 0.65 \\
Skill-only    & 117.6 & 0.49 & 0.80 \\
\textbf{SWATi [Proposed]}        & 167.4 & 0.62 & 0.90 \\
\bottomrule
\end{tabularx}
\vspace{-0.1in}
\end{table}

\vspace{-0.1in}
\section{Conclusion and Future Plan}

This paper presented a hybrid decentralized volunteer crowdsourcing framework integrating LLM-assisted semantic preprocessing, the SWATi skill- and willingness-aware assignment algorithm, and blockchain-based execution. In this architecture, LLMs normalize skills and preference cues from unstructured profiles, SWATi performs utility-driven volunteer–task matching, and blockchain ensures transparent and auditable execution.

By separating semantic extraction, algorithmic decision-making, and decentralized enforcement, the framework enables the use of expressive AI models while preserving deterministic and trustworthy task allocation. Preliminary results show that embedding SWATi within this layered architecture improves assignment utility and task coverage over random and skill-only baselines while maintaining predictable scalability.

Future work will investigate adaptive tuning of SWATi parameters, support collaborative multi-volunteer tasks with fairness constraints, and strengthen the integration between semantic preprocessing and decentralized execution through verifiable inference pipelines. Evaluation on live platforms will further enable analysis of volunteer engagement and long-term workforce sustainability.

Overall, SWATi provides an interpretable and efficient core for scalable and trustworthy volunteer crowdsourcing platforms.

\bibliographystyle{splncs04}
\bibliography{mybibliography}

\begin{thebibliography}{10}
\providecommand{\url}[1]{\texttt{#1}}
\providecommand{\urlprefix}{URL }
\providecommand{\doi}[1]{https://doi.org/#1}

\bibitem{snehaanbhawal_resume_dataset}
Anbhawal, S.: Resume dataset. \url{https://tinyurl.com/b54b3zhc} (2020), accessed: Jan 2026

\bibitem{borgave2023resume}
Borgave, S., Gavali, V., Kudumbale, S., Saoji, S.: Resume shortlisting and grading using tf-idf, cosine similarity and knn. Journal of Emerging Technologies and Innovative Research (JETIR)  \textbf{10}(5) (2023)

\bibitem{brown2020language}
Brown, T., Mann, B., Ryder, N., Subbiah, M., Kaplan, J.D., Dhariwal, P., Neelakantan, A., Shyam, P., Sastry, G., Askell, A., et~al.: Language models are few-shot learners. Advances in neural information processing systems  \textbf{33},  1877--1901 (2020)

\bibitem{cazzanti2006information}
Cazzanti, L., Gupta, M.R.: Information-theoretic and set-theoretic similarity. In: 2006 IEEE international symposium on information theory. pp. 1836--1840. IEEE (2006)

\bibitem{cerebrum2025decentralized}
{Cerebrum}: Decentralized compute network. \url{https://docs.cerebrum.dev} (2025), cerebrum Project Documentation

\bibitem{Chen2014}
Chen, Z., Fu, R., Zhao, Z., Liu, Z., Xia, L., Chen, L., Cheng, P., {Chen Cao}, C., Tong, Y., {Jason Zhang}, C.: {gMission: A general spatial crowdsourcing platform}. Proceedings of the VLDB Endowment  \textbf{7}(13),  1629--1632 (2014). \doi{10.14778/2733004.2733047}

\bibitem{Cheng2016}
Cheng, P., Lian, X., Chen, L., Han, J., Zhao, J.: Task assignment on multi-skill oriented spatial crowdsourcing. IEEE Transactions on Knowledge and Data Engineering  \textbf{28}(8),  2201--2215 (2016)

\bibitem{deerwester1990indexing}
Deerwester, S., Dumais, S.T., Furnas, G.W., Landauer, T.K., Harshman, R.: Indexing by latent semantic analysis. Journal of the American society for information science  \textbf{41}(6),  391--407 (1990)

\bibitem{feng2019mcs}
Feng, W., Yan, Z.: Mcs-chain: Decentralized and trustworthy mobile crowdsourcing based on blockchain. Future Generation Computer Systems  \textbf{95},  649--666 (2019)

\bibitem{goel2014}
Goel, G., Nikzad, A., Singla, A.: Allocating tasks to workers with matching constraints: truthful mechanisms for crowdsourcing markets. In: Proceedings of the 23rd International Conference on World Wide Web. pp. 279--280 (2014)

\bibitem{golem2016decentralized}
{Golem Project}: Golem: A decentralized computation marketplace. \url{https://golem.network/doc/Golemwhitepaper.pdf} (2016), white paper

\bibitem{hao2022structured}
Hao, Y., Sun, Y., Dong, L., Han, Z., Gu, Y., Wei, F.: Structured prompting: Scaling in-context learning to 1,000 examples. arXiv preprint arXiv:2212.06713  (2022)

\bibitem{jaccard1901distribution}
Jaccard, P.: Étude comparative de la distribution florale dans une portion des alpes et du jura. Bulletin de la Société Vaudoise des Sciences Naturelles  \textbf{37},  547--579 (1901), introduces the Jaccard similarity coefficient

\bibitem{9874795}
Kadadha, M., Singh, S., Mizouni, R., Otrok, H.: A context-aware blockchain-based crowdsourcing framework: Open challenges and opportunities. IEEE Access  \textbf{10},  93659--93673 (2022). \doi{10.1109/ACCESS.2022.3203850}

\bibitem{kushwaha2022ethereum}
Kushwaha, S.S., Joshi, S., Singh, D., Kaur, M., Lee, H.N.: Ethereum smart contract analysis tools: A systematic review. Ieee Access  \textbf{10},  57037--57062 (2022)

\bibitem{le2024metacrowd}
Le, H.D., Truong, V.T., Hoang, D.N., Nguyen, T.V., Le, L.B.: Metacrowd: Blockchain-empowered metaverse via decentralized machine learning crowdsourcing. In: 2024 IEEE Wireless Communications and Networking Conference (WCNC). pp.~1--6. IEEE (2024)

\bibitem{li2018crowdbc}
Li, M., Weng, J., Yang, A., Lu, W., Zhang, Y., Hou, L., Liu, J.N., Xiang, Y., Deng, R.H.: Crowdbc: A blockchain-based decentralized framework for crowdsourcing. IEEE transactions on parallel and distributed systems  \textbf{30}(6),  1251--1266 (2018)

\bibitem{li2022blockchain}
Li, S., Bai, X., Wei, S.: Blockchain-based crowdsourcing framework with distributed task assignment and solution verification. Security and Communication Networks  \textbf{2022}(1),  9464308 (2022)

\bibitem{Liu2016}
Liu, J., Zhu, H., Chen, X.: {Complicated-skills-based task assignment in spatial crowdsourcing}. Lecture Notes in Computer Science (including subseries Lecture Notes in Artificial Intelligence and Lecture Notes in Bioinformatics)  \textbf{9998 LNCS},  211--223 (2016). \doi{10.1007/978-3-319-47121-1_18}

\bibitem{nakamoto2008bitcoin}
Nakamoto, S.: Bitcoin: A peer-to-peer electronic cash system  (2008)

\bibitem{Ni2020}
Ni, W., Cheng, P., Chen, L., Lin, X.: {Task allocation in dependency-aware spatial crowdsourcing}. Proceedings - International Conference on Data Engineering  \textbf{2020-April},  985--996 (2020). \doi{10.1109/ICDE48307.2020.00090}

\bibitem{ouyang2022training}
Ouyang, L., Wu, J., Jiang, X., Almeida, D., Wainwright, C., Mishkin, P., Zhang, C., Agarwal, S., Slama, K., Ray, A., et~al.: Training language models to follow instructions with human feedback. Advances in neural information processing systems  \textbf{35},  27730--27744 (2022)

\bibitem{pazzani2007content}
Pazzani, M.J., Billsus, D.: Content-based recommendation systems. In: The adaptive web: methods and strategies of web personalization, pp. 325--341. Springer (2007)

\bibitem{filecoin2017}
{Protocol Labs}: Filecoin: A decentralized storage network. \url{https://filecoin.io/filecoin.pdf} (2017), accessed: Jan. 2026

\bibitem{samantafogirecruiter}
Samanta, R., Ghosh, S.K.: {FogiRecruiter : A fog-enabled selection mechanism of crowdsourcing for disaster management}. Concurrency and Computation: Practice and Experience pp.~1--9 (2022). \doi{https://doi.org/10.1002/cpe.7207}, \url{https://onlinelibrary.wiley.com/doi/abs/10.1002/cpe.7207}

\bibitem{samanta2024sustainable}
Samanta, R., Ghosh, S.K.: Sustainable volunteer engagement: Ensuring potential retention and skill diversity for balanced workforce composition in crowdsourcing paradigm. arXiv preprint arXiv:2408.11498  (2024)

\bibitem{samanta2021swill}
Samanta, R., Ghosh, S.K., Das, S.K.: Swill-tac: Skill-oriented dynamic task allocation with willingness for complex job in crowdsourcing. In: 2021 IEEE Global Communications Conference (GLOBECOM). pp.~1--6. IEEE (2021)

\bibitem{samanta2025enhancing}
Samanta, R., Ghosh, S.K., Das, S.K.: Enhancing crowdsourcing through skill and willingness-aligned task assignment with workforce composition balance. Pervasive and Mobile Computing  \textbf{107},  102012 (2025)

\bibitem{10925570}
Samanta, R., Ghosh, S.K., Das, S.K.: Sosta: Skill-oriented stable task assignment with bidirectional preferences in crowdsourcing. IEEE Transactions on Emerging Topics in Computing  \textbf{13}(3),  947--963 (2025). \doi{10.1109/TETC.2025.3548672}

\bibitem{Sama2212:Volunteer}
Samanta, R., Saxena, V., Ghosh, S., Das, S.K.: Volunteer selection in collaborative crowdsourcing with adaptive common working time slots. In: 2022 IEEE Global Communications Conference: Selected Areas in Communications: Social Networks (Globecom2022 SAC SN). Rio de Janeiro, Brazil (Dec 2022)

\bibitem{samanta2024empowering}
Samanta, R., Sethi, B., Ghosh, S.K.: Empowering volunteer crowdsourcing services: A serverless-assisted, skill and willingness aware task assignment approach for amicable volunteer involvement. arXiv preprint arXiv:2408.11510  (2024)

\bibitem{Song2020}
Song, T., Xu, K., Li, J., Li, Y., Tong, Y.: Multi-skill aware task assignment in real-time spatial crowdsourcing. GeoInformatica  \textbf{24},  153--173 (2020)

\bibitem{10126111}
Volpe, G., Mangini, A.M., Fanti, M.P.: A deep reinforcement learning approach for competitive task assignment in enterprise blockchain. IEEE Access  \textbf{11},  48236--48247 (2023). \doi{10.1109/ACCESS.2023.3276859}

\bibitem{wood2014ethereum}
Wood, G., et~al.: Ethereum: A secure decentralised generalised transaction ledger. Ethereum project yellow paper  \textbf{151}(2014),  1--32 (2014)

\bibitem{xu2022blockchain}
Xu, W., Duan, H., Chen, X., Huang, J., Liu, D., Chen, Y.: Blockchain-based multi-skill mobile crowdsourcing services. EURASIP Journal on Wireless Communications and Networking  \textbf{2022}(1), ~55 (2022)

\end{thebibliography}

\end{document}